\documentclass[aps,prl,amsmath,amssymbb,reprint,superscriptaddress]{revtex4-2}

\usepackage{graphicx}
\usepackage{dcolumn}
\usepackage{bm}

\usepackage[utf8]{inputenc}
\usepackage[T1]{fontenc}
\usepackage{graphicx}
\usepackage{enumitem}
\usepackage{amssymb}
\usepackage{hyperref}
\usepackage{amsthm}
\usepackage{cases}
\usepackage{tensor}
\usepackage{physics}
\usepackage{amsmath,amsfonts,amssymb}
\usepackage{color}
\hypersetup{
   colorlinks,
    citecolor=blue,
    filecolor=black,
    linkcolor=blue,
    urlcolor=black
}

\definecolor{darkorange}{rgb}{1.0, 0.55, 0.0}

\newcommand{\sm}{ \! - \!}

\newcommand{\Itot}{ \mathcal{I}_{\rm tot}}

\usepackage[normalem]{ulem}

\begin{document}

\title{Epidemic models on homogeneous networks : some analytical results}

\newcommand{\majulab}{MajuLab, CNRS–UCA-SU-NUS-NTU International Joint Research Laboratory}
\newcommand{\cqt}{Centre for Quantum Technologies, National University of Singapore, 117543 Singapore, Singapore}
\newcommand{\LPTMS}{Université Paris-Saclay, CNRS, LPTMS, 91405, Orsay, France}
\newcommand{\PCS}{Center for Theoretical Physics of Complex Systems, Institute for Basic Science, Daejeon 34126, Korea}

\author{Louis Bremaud}
\affiliation{\LPTMS}
\author{Olivier Giraud}
\affiliation{\LPTMS}
\affiliation{\majulab}
\affiliation{\cqt}
\author{Denis Ullmo}
\affiliation{\LPTMS}

\date{\today}

\begin{abstract}

The ability to actually implement epidemic models is a crucial stake for public institutions, as they may be overtaken by the increasing complexity of current models and sometimes tend to revert to less elaborate models such as the SIR. In our work, we study a simple epidemic propagation model, called SIR-$k$, which is based on a homogeneous network of degree $k$, where each individual has the same number $k$ of neighbors. This model is more refined than the basic SIR which assumes a completely homogeneous population.  We show that nevertheless,  analytical expressions, simpler and richer than the ones existing for the SIR model, can be derived for this SIR-$k$ model. In particular we obtain an exact implicit analytical solution for any $k$, from which quantities such as the epidemic threshold or the total number of agents infected during the epidemic can be obtained.  We furthermore obtain simple exact explicit solutions for small $k$'s,  and in the large $k$ limit we find a new formulation of the analytical solution of the basic SIR model, which comes with new insights.

\end{abstract}
\maketitle

Understanding the dynamics of epidemics is of primary importance to allow public policies to mitigate  their negative impact \cite{toda2020susceptible, social_impact_covid}. Models of epidemic propagation have therefore been introduced as early as one century ago, in 1927, with in particular the seminal paper of Kermack and McKendrick \cite{1927_kermack}. In this paper, they introduce the SIR model, 
which, despite its simplicity, is still a basis of work in many studies \cite{mathematics_infectious_diseases, anderson1991infectious, Covid_France_science}. This model divides a population into susceptible, infected and recovered individuals, and two parameters characterize the evolution: the transmission rate $\beta$ and the recovery rate $\gamma$. In the simplest version of the model, $\beta$ and $\gamma$ are assumed to be constant on the epidemic time scale. The time evolution of the fractions $(S, I, R)$ of susceptible, infected, and recovered agents is then given \cite{mathematical_theory_of_infectious_disease, anderson1991infectious} by
\begin{equation}
\label{eq:SIR_simple}
\begin{aligned}
{\dot S} &= - \beta S I \\
{\dot I} &= \beta S I - \gamma I \\
{\dot R} &= \gamma I\;.
\end{aligned}
\end{equation}
This system of differential equations has been studied in detail during the past century \cite{Kendall_1956, mathematical_theory_of_infectious_disease, anderson1991infectious}; in particular, explicit solutions describing the beginning of epidemics \cite{1927_kermack}, and complete implicit solutions \cite{Exact_analytical_SIR_2014,Prodanov2022_analytical_SIR,carvalho2021analytical}, have been derived.

Even though the basic SIR model has been successful, it can be considered too simplistic. This is why more accurate variants \cite{SIRC, SEIR, SIRD, SIRV, Asymptotic_SIR_SEIR} and a number of more complex models \cite{Inferring_social_structure,mistry2020inferring, Impact_NPI_Imperial, Bremaud_Ullmo_PhysRevE, mathematical_theory_of_infectious_disease} have been since then introduced. Among these models, SIR models on networks provide a good balance between simplicity, physical understanding, and improved accuracy \cite{Watts1998-small_world, PhysRevE_cooperation_small_world_networks, PhysRevE_pair_quenched_MFT_threshold_SIS, PhysRevE_synergistic_epidemic, Barthelemy2005_networks, dynamics_spreading_homogeneous_heterogeneous, Asymptotic_SIR_SEIR, Epidemc_spread_networks,newman2002spread}. This approach benefitted both from the wealth of activity in network theory in the past two decades and from the increased availability of large amounts of data \cite{didomenico_thesis} about contact networks (see \cite{Epidemic_processes_complex_networks,Wang_2017_unification} for a complete review on the subject).   

In spite of their success in extending the basic SIR model to more complex and accurate frameworks, these network models so far lack one important feature which is the existence of analytical results for the solution of the models' equations.  Our goal in this paper is to provide such analytical results in the case of homogeneous networks, which are characterized by their constant connectivity $k$. For any  given value of $k$ we obtain analytic expressions analogous to (and in some circumstances stronger than) the ones existing for the SIR model \eqref{eq:SIR_simple};  when $k=2$ or $3$ we obtain simple explicit expressions, while in the limit $k\to\infty$ we recover the limiting case of the basic SIR, leading to some new physical insight as well as some useful approximations of this well-known model.

\paragraph{SIR model on a homogeneous network with k neighbors}
We consider a population of individuals which can be in one of the three possible states (susceptible, infected, recovered). Each agent is in contact  with $k$ fixed neighbors only; the standard SIR model, where everyone is in contact with everyone, corresponds to the large-$k$ limit of our model. The population can be represented by a homogeneous network with fixed connectivity $k$, where each node corresponds to an individual and edges connect neighboring individuals. Associated with each of these edges is a probability $\lambda dt$ that an infected individual will infect a (susceptible) neighbor during the time interval $[t,t+dt]$. As in the basic SIR model, infected individuals may also recover from the disease  during that time interval with a probability $\gamma dt$. The epidemic then spreads through the network following  a standard Markovian process (see \cite{stochastic_models} for a detailed procedure), and dynamic quantities are averaged over realizations of the network and of the Markovian process.

\begin{figure}[!t]
\centering
\includegraphics[width=\linewidth]{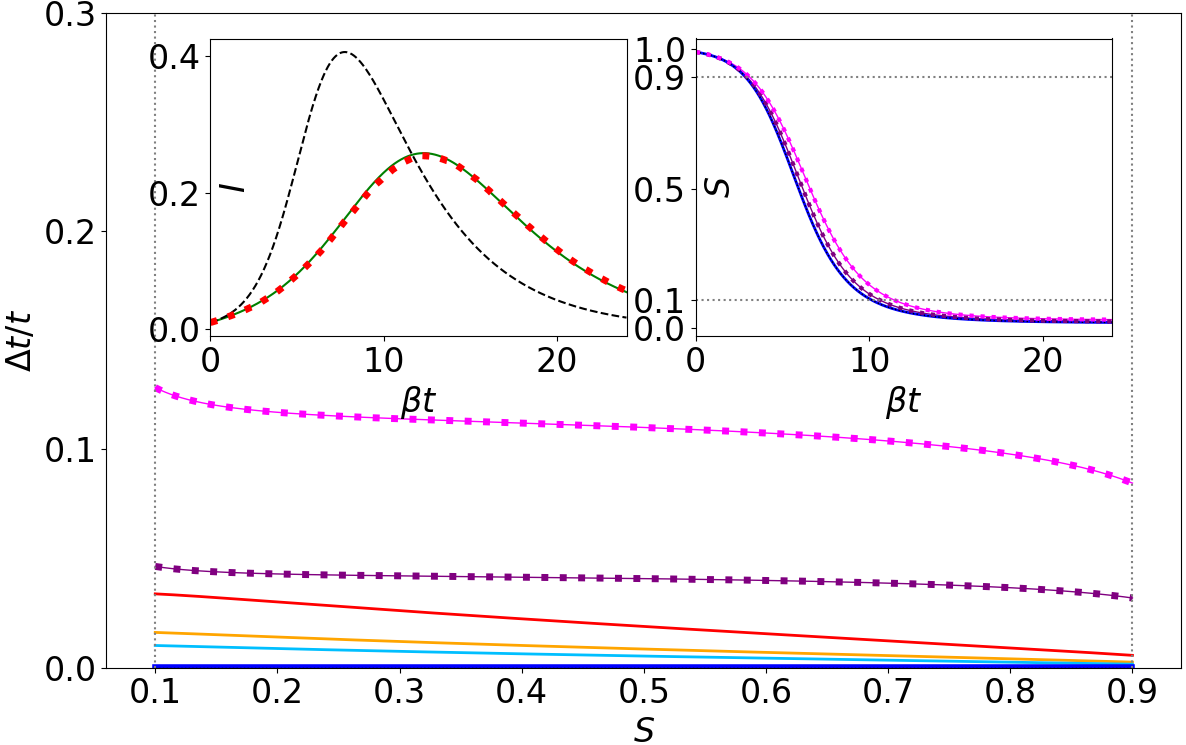}
\caption{
Main panel: Time delay $\Delta t=t(S)-t_{\textrm{SIR}}(S)$ with $t_{\textrm{SIR}}$ obtained by numerically solving \eqref{eq:SIR_simple}. Solid thick dark blue: analytical expression \eqref{eq:t_S_Lambert}, corresponding to the limit case  SIR-${\infty}$.  Purple ($k=50$) and magenta ($k=20$) plots:  numerical resolution of the SIR-$k$ model \eqref{eq:SIR_k1} (solid lines) and corresponding analytical solution \eqref{eq:tdeS} (dots). Solid red, orange, light blue :  approximation  \eqref{eq:t_S_Lambert_Nj_2} with $m=2,3,4$ respectively.  
Right inset: proportion of susceptible $S(t)$ for  the same configurations (except for the approximations \eqref{eq:t_S_Lambert_Nj_2} which are indistinguishable from the exact SIR). The gray horizontal dotted lines indicate the range of $S$  values taken for the main panel. 
Left inset: proportion of infected $I(t)$ for $k=5$.  
Red dotted line: numerical resolution of the SIR-5 model Eqs.~\eqref{eq:SIR_k1}-\eqref{eq:SIR_k2}  ; green solid line: average over $100$ realizations of the Markovian process of an epidemic on a large homogeneous network of degree $k=5$, with $N=3000$ nodes (with random initial infected nodes);  black dashed line:  basic SIR model with $\beta = \lambda k$ . For the parameters, we take $\mu = 0.25$, $S_0=0.99$.} 
\label{Fig:S_comparison}
\end{figure}

The time evolution of the average fractions $S(t)$, $I(t)$, and $R(t)$ of susceptible, infected and recovered individuals requires to take into account correlations between the states of two neighbors, which are very strong in a network. For a SIR model on a $k$-homogeneous network we obtain the system of equations 
\begin{subequations}
\label{eq:SIR_k1}
\begin{align}
{\dot S} &= - \lambda k G^{si} S  \label{eq:SIR_k1a} \\
{\dot I} &= \lambda k G^{si} S - \gamma I  \label{eq:SIR_k1b}\\
{\dot R} &= \gamma I \label{eq:SIR_k1c}\; , 
\end{align}
\end{subequations}
with $S(t)+I(t)+R(t) = 1$. Here $G^{si}(t)$ corresponds to the probability that a neighbor of a given susceptible individual is itself infected; thus $k G^{si}(t)$ is the average number of infected individuals in the neighborhood of a susceptible individual. Introducing  $G^{ss}(t)$ and $G^{sr}(t)$ in a similar way, with  $G^{ss}(t) + G^{si}(t) + G^{sr}(t) = 1$, the time dependence of these two-point correlators is given by 
\begin{subequations}
\label{eq:SIR_k2}
\begin{align}
   \dot{[SG^{ss}]} &= - 2 S G^{ss} (k-1) G^{si} \lambda    \label{eq:SIR_k2a} \\
    \dot{[SG^{si}]} &= S G^{ss} (k \sm 1) G^{si} \lambda  \label{eq:SIR_k2b}\\
    & \quad - S G^{si} \left[ (k\sm 1)G^{si} + 1 \right] \lambda - \gamma SG^{si} \notag \\
    \dot{[SG^{sr}]} &= \gamma SG^{si} - SG^{sr} (k-1)G^{si} \lambda \;.
    \label{eq:SIR_k2c} 
\end{align}
\end{subequations} 
To derive \eqref{eq:SIR_k2}  we made the pairwise approximation \cite{compact_pairwise_heteregogeneous}, that is, we neglected three-point correlations (and beyond) which should appear in the evolution of $G^{si}$.  This approximation has been used in  \cite{Wang_2017_unification} to derive equations for the SI model on a generic network. Furthermore, in the case of  homogeneous networks  with a large number of nodes $N \to \infty$, as we consider here, the fraction of loops with arbitrary finite size vanishes \cite{wormald1981asymptotic, dorogovtsev2003spectra, goirand2021network}, and therefore the correlations beyond two-point ones can be neglected.   Equations~\eqref{eq:SIR_k1}-\eqref{eq:SIR_k2} form what we will  call the ``SIR-$k$ model'' in the following. In Fig.~\ref{Fig:S_comparison} (left inset), we demonstrate the accuracy of our approximation by comparing a numerical solution of Eqs.~\eqref{eq:SIR_k1}-\eqref{eq:SIR_k2} with a Markovian evolution of a population according to the same dynamics. The parameters of our problem are $S_0$ the initial proportion of susceptible agents, $k$ the number of neighbors,  $\beta= \lambda k$ the contagiousness and $\gamma$ the recovery rate, which leads to a dimensionless quantity $\mu = \gamma/\beta$ driving the epidemic, while $\beta$ only changes the time scale (see for example \cite{Prodanov2022_analytical_SIR}).


\paragraph{Analytical solution of the SIR-$k$ equations}
From Eqs~\eqref{eq:SIR_k1}-\eqref{eq:SIR_k2}, we can obtain an ordinary differential equation involving only $S(t)$. Inserting $G^{si} = - \dot S/(\beta S)$, which we get from  Eq.~\eqref{eq:SIR_k1a}, into Eq.~\eqref{eq:SIR_k2a}, we have  
\begin{equation} \label{eq:SIR_k2bis}
\frac{\dot{[SG^{ss}]}}{S G^{ss}} = 2 \frac{k-1}{k} \frac{\dot S}{S} \; .
\end{equation}
At $t=0$,  $S(0) = S_0=G^{ss}(0)$ if we assume that there are no correlations at time $0$ (i.e.~the neighborhood of infected and susceptible individuals is the same). Then Eq.~\eqref{eq:SIR_k2bis} can be integrated as $G^{ss} = S_0^{\frac{2}{k}} S^{1 - \frac{2}{k}}$.  Using Eq.~\eqref{eq:SIR_k1a} and this expression for $G^{ss}$, Eq.~\eqref{eq:SIR_k2b} yields
\begin{equation}
\label{eq:Sdotdot}
\ddot{S} = \lambda S_0^{1-\frac{2}{k}} S^{1 - \frac{2}{k}} (k-1) \dot{S}  + \frac{k-1}{k} \frac{\dot{S}^2}{S} - (\gamma + \lambda) \dot{S}.
\end{equation}
This is a second-order differential equation in $S$ that we need to integrate twice. A first integration is obtained by dividing \eqref{eq:Sdotdot} by $\dot{S}$ and introducing  $\varphi(S) = \dot{S}$, which verifies  
\begin{equation}
\label{eq:ED_phi_S}
\frac{d\varphi(S)}{dS} =\lambda S_0^{\frac{2}{k}} S^{1 - \frac{2}{k}} (k-1)  + \frac{k-1}{k} \frac{\varphi(S)^2}{S} - (\gamma + \lambda) \;.
\end{equation}
Equation \eqref{eq:ED_phi_S} can be integrated as an equation in the variable $S$. Changing to the variable $z \equiv (S/S_0)^{\frac{1}{k}}$,  and using $\mu = \gamma/ \beta$, this gives
\begin{equation}
\label{eq:P(z)}
 \dot{z} = \lambda P(z) \; , \qquad P(z)= S_0 z^{k-1} - (k\mu+1)z + k \mu \; .
\end{equation}
Separating the variables $z$ and $t$ and using the partial fraction decomposition of $1/P(z)$ in terms of the roots $z_j$ $(j=0,\cdots, k \sm 2)$ of $P(z)$, the integral of  Eq.~\eqref{eq:P(z)} becomes
\begin{equation}
\label{eq:solve_y_1}
\int_{1}^{z} {\frac{dz'}{P(z')}} = 
\sum_{j=0}^{k-2}  \int_{1}^{z}  {\frac{A_j} {z' - z_j}} dz'= \lambda t \; ,
\end{equation}
with 
\begin{equation}
\label{eq:defAi}
A_j=\frac{1}{P'(z_j)}=\frac{1}{\prod_{l\neq j}(z_j-z_l)}\;.
\end{equation}
Equation \eqref{eq:solve_y_1} readily gives an  explicit expression for $t$ as a function of $S$ as
\begin{equation}
\label{eq:tdeS}
t(S) =\frac{1}{\lambda}\sum_{j=0}^{k-2} A_j \log \left( \frac{(S/S_0)^{1/k} - z_j}{1-z_j}  \right)   \; .
\end{equation}
Note that the complex roots $z_j$  are pairwise complex conjugate so that the whole sum is real, as it should. One then gets a parametric solution for the number of infected individuals under the form $(t(S),I(S))$ by integrating \eqref{eq:SIR_k1b}. We checked, for many different values of the parameters  ($S_0$, $\mu$, $k$), that the analytical solution \eqref{eq:tdeS} perfectly reproduces the numerical resolution of \eqref{eq:SIR_k1}-\eqref{eq:SIR_k2},  and we illustrate it for one example in Fig.~\ref{Fig:S_comparison}.
Note that a similar approach allows to address the SI model, which corresponds to the limit $\mu \to 0$; in that case we get
\begin{equation}
\label{eq:S_SI}
    S(t) = S_0^{-\frac{2}{k-2}} \left( \frac{1 - S_0}{S_0} e^{\lambda(k-2)t} +1 \right)^{-\frac{k}{k-2}} \; ,
\end{equation}
which in the limit $k \to \infty$ coincides with the known solution of the SI model \cite{mathematical_theory_of_infectious_disease}.


\begin{figure}[!t]
\centering
\includegraphics[width=\linewidth]{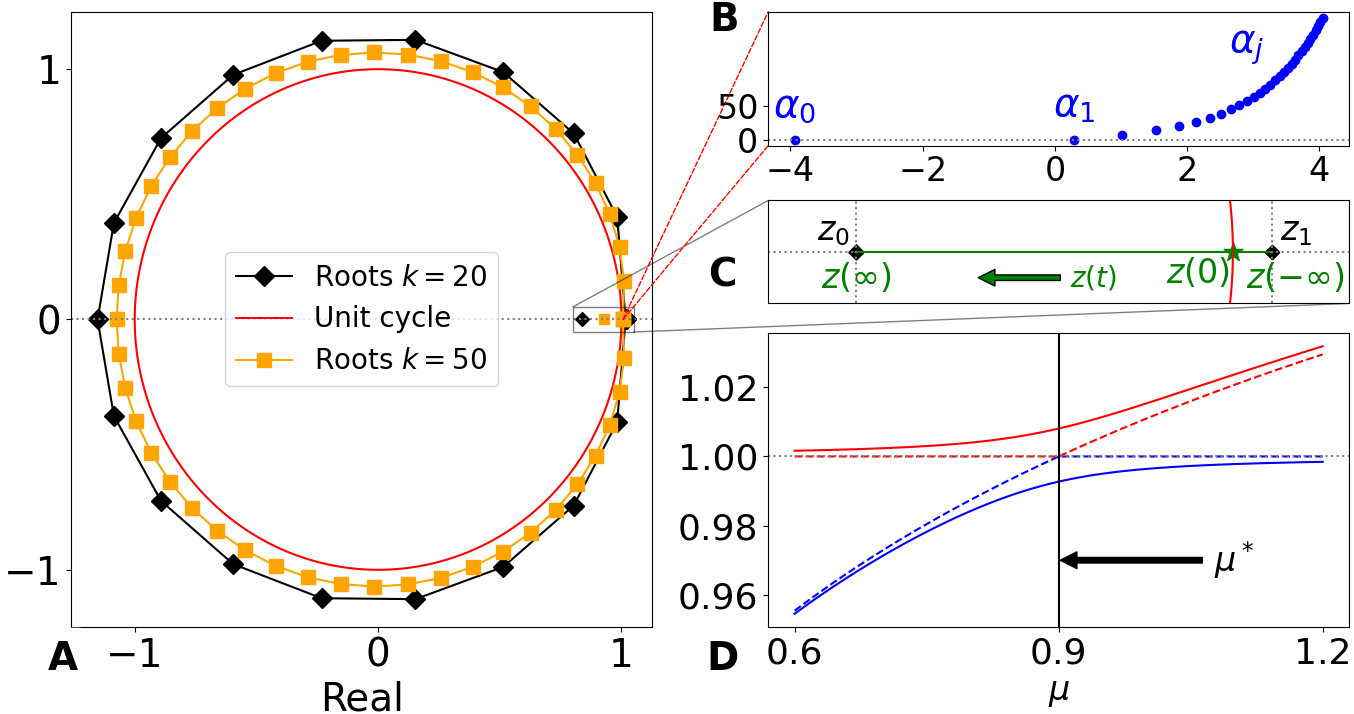}
\caption{\textbf{A}. Orange squares (resp.~black diamonds): location, in the complex plane, of the roots of the polynomial $P(z)$  Eq.~\eqref{eq:P(z)} for $k=50$  (resp.~$k=20$) with $S_0= 0.8$ and $\mu = 0.25$ 
\textbf{B} Blow-up showing, in the complex plane, the limit as $k\to\infty$ of the $\alpha_j$ defined by $z_j = 1 + \alpha_j/k$.  \textbf{C} Zoom on the complex plane close to $1$ with $z(t) = (S(t)/S_0)^{1/k}$ traveling the green line from $z_1 = z({-\infty})$ to $z_0 = z({\infty})$  and passing  through $z(0)=1$.
\textbf{D}  Blue line (resp. red line): illustration, for  $k=20$, of the variation with $\mu$ of the roots $z_0(\mu)$ (resp.~$z_1(\mu)$) for $S_0 = 0.99$ (solid line) and $S_0=1$ (dashed line). The value $\mu_k^*$ such that $z_0(\mu^*_k) = z_1(\mu^*_k) = 1$ is the epidemic threshold.}
\label{Fig:roots}
\end{figure}

We now comment on the consequences of Eq.~\eqref{eq:tdeS}.
Polynomials such as $P(z)$ in Eq.~\eqref{eq:P(z)} have a long history, dating back to Lambert \cite{lambert1758observationes, On_LambertW_function} and Euler \cite{euler1783serie}.  In particular, one can explicitly express all the roots $z_j$ of $P(z)$  as infinite series in $a = S_0^{-1/k}k\mu b^{(1-k)/(k-2)}$ with $b =(k\mu+1)S_0^{-2/k}$ \cite{belkic2019all, glasser2000hypergeometric}. As illustrated in Fig.~\ref{Fig:roots}A, for $k>2$ there are two real positive roots, $z_{0}\in [0,1]$ and  $z_1  \in [1,\infty[$. 
Since  $S/S_0 \in [0,1]$,  the only possible divergence of $t$  in \eqref{eq:tdeS} corresponds to the root $z_0$, and we thus get that $S_\infty \equiv {\lim}_{t \to \infty}S(t) = S_0 z_0^k$. 
A useful quantity for  public agencies in charge of controlling the epidemic  (see \cite{Kendall_1956} for the basic SIR model) is the fraction of the population which will be infected during the course of the epidemic; it can be expressed as $ \Itot^{(k)} = S_0 - S_\infty = S_0 (1 - z_0^k)$. The second positive real root $z_1$ can then be interpreted as the non-physical limit  to which $S$ would tend if one follows the SIR-$k$ equations for negative times, $S_{-\infty} \equiv {\lim}_{t \to -\infty} S(t) = S_0 z_1^k>1$.  As illustrated in Fig.~\ref{Fig:roots}C, the associated quantity $z(t) = (S(t)/S_0)^{1/k}$ decreases from 1 to $z_0$  for $t\in[0, + \infty[$, and from $z_1$ to 1 for the non-physical part $t\in ]-\infty,0]$.

Whatever the value of $\mu$ and $k$,  $P(1) = S_0 - 1$. Thus, as illustrated in Fig.~\ref{Fig:roots}D, $z=1$ cannot be a root of $P(z)$ for $S_0<1$, but always is for $S_0 = 1$. In this latter case, two situations can occur : either $z_1 =1$ and $z_0 < 1$, in which case an epidemic starting with  $S_0=1$ (i.e.~with an infinitesimal fraction of infected individuals) will eventually propagate into the network and infect a finite fraction of the population ; or $z_0 = 1$ and $z_1 \ge 1$, and thus $S_\infty = 1$: an epidemic starting with  $S_0=1$ does not propagate. The value $\mu^*_k$ of the parameter $\mu$ corresponding to the transition between these two regimes is the threshold beyond which, for $S_0 = 1$, the epidemic does not spread. At the threshold, $z=1$ is a double root of $P(z)$ and thus $\mu^*_k = ({k-2})/{k}$. As $k\to\infty$ we get $\mu^*_k\to 1$, which coincides with the result of Kermack and McKendrick \cite{1927_kermack} for the original SIR model.

\paragraph{Small number of neighbors}
It is possible to invert the expression \eqref{eq:tdeS} for $k = 2$ and 3. Consider first the case $k=2$. A connected network then corresponds to a single loop of size $N$. There is only one root $z_0 =2 \mu /({I_0 + 2 \mu })$,  with $I_0= 1-S_0$ the initial fraction of infected individuals. We can therefore write \eqref{eq:tdeS} as 
\begin{equation}
\label{eq:tdeS_k2}
t(S)=\frac{1}{\lambda} A_0 \log \left( \frac{(S/S_0)^{1/2} - z_0}{1-z_0}  \right)   \; ,
\end{equation}
with $A_0 = -1/(I_0 + 2\mu)  < 0 $.  Inverting  \eqref{eq:tdeS_k2} we get an explicit solution for the SIR-2 model as \begin{equation}
    \label{eq:Sdet_k2}
S(t) = S_0 \left[ 1 + \frac{I_0\left( e^{-t/\tau}-1 \right) }{I_0+2\mu} \right]^2 , \quad
\tau  = \frac{1}{\lambda(2\mu + I_0)} \; .
\end{equation} $S(t)$  thus follows an exponential decay with rate $\tau$ and converges to $S_{\infty} = S_0 z_0^2$, as expected. We get $\Itot^{(2)} = S_0 \left(1- (1-{I_0}/({2\mu}))^{-2} \right)$, which varies from $S_0$ for strong epidemic ${I_0}/{\mu} \gg 1$ to $0$ with ${I_0}/{\mu}\ll 1$. In particular  $\lim_{S_0 \to 1} \Itot^{(2)} = 0$ for any positive value of $\mu$, which can also be seen from the fact that $\mu^*_2= (k-2)/k = 0$. This is unique to the $k=2$ case because of its essentially $1$d geometry, which implies that the number of infected agents caused by a single ``patient zero'' is necessarily finite.

Turning now to the case $k=3$, we get $P(z) = S_0 z^2 - (3\mu+1) z + 3 \mu $, which has two (real positive) roots
\begin{equation}
\label{eq:z_k3}
z_{0,1} = \frac{1}{2 S_0} \left[{(3\mu + 1)} \pm {\sqrt{(3\mu+1)^2-12\mu S_0}} \right] \; ,
\end{equation}
yielding
\begin{equation}
\label{eq:tdeS_k3}
t(S)= \frac{A_0}{\lambda} \log \left[ \frac{\left((S/S_0)^{1/3} - z_0\right) (1-z_1)}{\left( (S/S_0)^{1/3} - z_1 \right) (1-z_0)} \right]   \; ,
\end{equation}
where we have used that $A_1 = -A_0 = 1/(z_1 - z_0)$. We can invert this formula to get 
\begin{equation}
\label{eq:Sdet_k3}
S(t)= S_0 \left( \frac{z_0 - z_1 B e^{{\lambda}{(z_0-z_1)}t}}{1 - Be^{\lambda(z_0-z_1)t}} \right)^3 ,\quad B = \frac{1-z_0}{1-z_1}.
\end{equation}
As expected, this expression verifies that  $S(0)=S_0$ and 
$S_{\infty} = S_0 z_0^3 $. The explicit expression for $\Itot^{(3)}$ is $S_0 - \frac{1}{8 S_0^2} \left [ (3\mu + 1)  + \sqrt{(3\mu+1)^2-12\mu S_0} \right ]^3$. 
For $S_0 = 1$, Eq.~\eqref{eq:z_k3} simplifies to $z_0 = \min(1,3\mu)$, $z_1 = \max(1,3\mu)$, and we recover $\mu^*_3 =\frac13$; for $\mu < \mu^*_3$,  $\Itot^{(3)} =1-(3 \mu)^3$, while for $\mu \ge \mu^*_3$ the epidemic does not propagate as $S_{\infty} = 1$.

Finally, we consider the case $k=4$, but limiting ourselves for simplicity to the limit $S_0 \to 1$ and the regime $\mu < \mu^*_4 =1/2$.  In that case $P(z)$  has three roots, which, introducing $\kappa = \sqrt{1/4+ 4\mu}$, can be written as  $z_0 = \kappa -\frac12, z_1 = 1, z_2 = - \kappa-\frac12$ with furthermore $A_0 = [\kappa(2\kappa + 3) ]^{-1},  A_1 = [2-4\mu]^{-1}, A_2 = [\kappa(2\kappa - 3) ]^{-1}$.  Introducing the time $t_0$ defined by 
\begin{equation}
\label{eq:t0}
t_0 = - \frac{1}{\lambda}\sum_{j=0}^{k-2} A_j \log \left| {z_j - 1}  \right|  
\; \underset{S_0  \to 1}{\sim} \;\frac{\log (1-S_0)}{\lambda (2+k(\mu-1))}   \; ,
\end{equation}
we see that $\lim_{S_0 \to 1} t_0 = \infty $ since in that case $z_1$ goes to 1.  This expresses the fact that the beginning of the epidemic takes an infinite amount of time as the initial proportion of infected individuals goes to zero, which is true for any $k>2$.   Scaling out this infinity we get
\begin{equation}
    t -t_0 = \frac{1}{\kappa \lambda } \sum_{\epsilon = \pm 1}
        \left(\frac{1}{2\kappa + 3\epsilon} \log \left|\frac{S^{1/k} + \epsilon \kappa + \frac12 }{ S^{1/k} -1}  \right| 
     \right)
\end{equation}
and $\Itot^{(4)} = (-16\mu^2 -8\mu +1/2) + (1+8\mu)\sqrt{4\mu + 1/4}$ (which is indeed such that $\Itot^{(4)}(\mu^*_4 )=0$).

\paragraph{Large-$k$ limit of the SIR-$k$ model}

Another interesting limit of the SIR-$k$ model is $k\to\infty$, through which we recover the original SIR model, but with a new point of view.  As illustrated in Fig.~\ref{Fig:roots}A,  $z_0$ and $z_1$ converge to 1 (from below and from above respectively) and all the other roots converge to the unit circle in the complex plane. This can be understood from their series expansion in \cite{belkic2019all, glasser2000hypergeometric}.  Using that $z_j$ is a root of $P(z)$ we can write the factor $A_j$ defined in  Eq.~\eqref{eq:defAi} as
\begin{equation}
    A_j =\left[(k-1) k \mu \frac{z_j-1}{z_j} - k (\mu-1)\right]^{-1} \; \; .
\end{equation}
For most roots of $P(z)$, $z_j - 1= O(k^0)$ (we refer to them as ``far from one'') and thus $A_j = O(k^{-2})$. It is only for the roots close to one, and more precisely such that $z_j - 1 = O(k^{-1})$, that $A_j = O(k^{-1})$. 
In the same way, the logarithm factors are $O(k^{-1})$ for the roots far from one and $O(k^0)$ for the roots close to one.  In Eq.~\eqref{eq:tdeS}, noting that $\lambda^{-1} = k \beta^{-1}$, we see that the sum over roots far from one involves $O(k)$ terms of order $O(k^{-2})$ and has therefore a negligible $O(k^{-1})$ contribution, whereas each root close to one has an $O(k^0)$ contribution.
We can thus write all relevant roots as $z_j = 1 +  \alpha_j/k$  where   $\alpha_j$  reaches a constant value as  $k\to\infty$.   Writing that $z_j$ is  a root of $P(z)$ thus reads
\begin{equation}
\label{eq:betaj_k}
    S_0 \left(1+\frac{\alpha_j}{k}\right)^{k-1} = k\mu \left[\left(1 + \frac{1}{k\mu}\right)\left(1+\frac{\alpha_j}{k}\right) -1 \right]
\end{equation}
which, taking the limit  $k\to\infty$ on both sides (with $\alpha_j$ now corresponding to that limit), gives $\exp({ \alpha_j}) = ({\mu}/{S_0}) \left(  {1}/{\mu}+ {\alpha_j} \right)$. Defining now $\gamma_j = {\alpha_j} + {1}/{\mu}$ and $\chi = (S_0/\mu)e^{-1/\mu}$, we get 
\begin{equation} \label{eq:gamma}
\chi = \gamma_j \exp(-\gamma_j) \; .
\end{equation}
Equation~\eqref{eq:gamma} can be rewritten in terms of the Euler $T$  function (see \cite{On_LambertW_function} for mathematical details) as $\gamma_j = T(\chi)$.  The $T$ function has two real branches $T_0$ and $T_{-1}$ which correspond to the two positive real roots of $P(z)$, and an infinite number of complex branches corresponding to the complex numbers $\gamma_j$. In particular we get for the first root $\lim_{k \to  \infty} S_\infty = \mu T_0(\chi)$, which is equivalent to the well-known self-consistent equation $S_\infty = 1 + \mu \ln({S_\infty}/{S_0})$ given for instance in \cite{mathematics_infectious_diseases}. In the large-$k$ limit we finally get 
\begin{align}
t(S) & = \frac{1}{\mu \beta} \sum_{j = 0}^{+ \infty} \frac{1}{\alpha_j + 1/\mu-1 } \log \left(1 - \frac{\log(S/S_0)}{ \alpha_j}\right)  \; , \nonumber\\
\alpha_j &= T_{-j}(\chi) - 1/\mu \; ,
\label{eq:t_S_Lambert}
\end{align}
In Fig.~\ref{Fig:S_comparison} we check the accuracy of this expression. 

An implicit analytical solution $t(S)$ for the SIR model \eqref{eq:SIR_simple} is known in the literature and takes the form of an integral (see for instance \cite{Exact_analytical_SIR_2014}). Our formula \eqref{eq:t_S_Lambert} is an alternative expression for  $t(S)$ and comes with interesting new insights, as it depends on quantities $\alpha_j$ which have an explicit expression. In Fig.~\ref{Fig:roots}B we show the first terms of the sequence. We see that $\alpha_0 < 0$ and $\alpha_1 > 0$ are indeed the two unique real values, while the subsequent $\alpha_j$ are purely complex; the latter are well approximated by $\alpha_j \simeq 2 \pi i j$ for large $j$ as the roots $z_j$ converge to the unit circle $\exp \left(\frac{2\pi i j}{k-2} \right)$. This allows us to give an accurate approximation for $t(S)$ with the first $m$ terms computed explicitly and the sum from $m$ to $\infty$ replaced by an integral that we evaluate, giving (for arbitrary $m$)
\begin{equation}
\label{eq:t_S_Lambert_Nj_2}
t(S) \simeq \frac{1}{\beta \mu} \sum_{j = 0}^{m-1} \frac{\log \left(1 - \frac{1}{ \alpha_j}\log(S/S_0)\right)}{\alpha_j + 1/\mu-1 } + \frac{1}{m}\frac{\log(S/S_0)}{ (2 \pi)^2 \beta \mu} \; .
\end{equation}
This simple expression of $t(S)$ is shown in Fig.~\ref{Fig:S_comparison} for $m = 2, 3, 4$.   We see that $m=2$, which only involves the two real roots $z_{0,1}$,  does already better than the SIR-$k$ model with $k=50$ to approximate the original SIR model. Furthermore, one can write a useful approximation in the limit of small $\mu$ and $S_0$ close to 1. Indeed, using  the Taylor expansion of $T_0(\chi)$ given in \cite{On_LambertW_function} and Eq.~\eqref{eq:betaj_k}, we get $\alpha_0 \simeq \chi + \chi^2 - 1/\mu  $  and $\alpha_1 \simeq (1-S_0)/(S_0-\mu)$. We checked that Eq.~\eqref{eq:t_S_Lambert_Nj_2} with $m=2$ and these values for $\alpha_{0,1}$ is numerically accurate for $S_0 > 0.95$ and $\mu < 0.5$, providing a compact and explicit analytical result, in a regime which corresponds to most of the practical use. Note that the $1/m$ correction in the r.h.s.\ of \eqref{eq:t_S_Lambert_Nj_2} is in practice fairly small even for low $m$.
 
\paragraph{Conclusion}

In this work, we have derived Eqs.~\eqref{eq:SIR_k1}-\eqref{eq:SIR_k2} for the SIR-$k$ model, and obtained an exact implicit expression of $t(S)$ \eqref{eq:tdeS},  valid for arbitrary $k$, as a finite sum over the roots $z_j$ of the polynomial $P(z)$ \eqref{eq:P(z)}. It turns out that the main qualitative properties of the epidemic dynamics are governed by its two positive real roots $(z_0,z_1)$. In particular the  proportion of agents infected during the total duration of the epidemic is given by $\Itot^{(k)} = S_0 (1 - z_0^k)$, and the threshold value of $\mu$ for which, even for an infinitely small initial proportion of  infected individuals,  a epidemic starts to propagate and affect a finite proportion of the agents,  is given by the condition $z_0(\mu^*_k) = z_1(\mu^*_k)=1$, leading to $\mu^*_k = (k-2)/k$.  This value is lower than its counterpart for the basic SIR model $\mu^*_{\rm SIR} = \mu^*_\infty = 1$,  which indicates that the  propagation of epidemics is more difficult in the SIR-$k$ model that in the basic SIR one. This is in contrast with heterogeneous networks, for which an epidemic spreads more easily  than in the SIR model.  Many other quantities of interest can be accessed via our approach, such as the time of the epidemic peak $t(I_{\max})$. In the cases $k=2$ and $k=3$ we get exact explicit expressions for $S(t)$. In the limit $k \to \infty$ we obtain new exact expressions for the original SIR model, together with useful approximate results that work extremely well numerically.  

The SIR-$k$ model on homogeneous networks thus presumably provides a good balance between increase of complexity and increase of effectiveness. It is characterized by only three parameters $(S_0, \mu, k)$ which, compared to the basic SIR, only adds parameter $k$ corresponding to the average number of possible contacts of individuals, a relatively accessible quantity in practice.  
In contrast to heterogeneous networks, which, despite their more refined description of the contact structure,  are not used systematically because of their complexity, our SIR-$k$ model is almost as simple as the basic SIR model. Indeed, it enjoys an exact solution simpler than the SIR one, while the numerical resolution is still fast and tractable ($6$ equations instead of $3$). We therefore hope that our work will encourage institutions to use the SIR-$k$ model in practice, instead of the basic SIR, all the more since the two yield significantly different outcomes when the number of neighbors if low, as shown in Fig.~\ref{Fig:S_comparison} left inset.

\bibliography{biblio_file}

\begin{thebibliography}{41}%
\makeatletter
\providecommand \@ifxundefined [1]{%
 \@ifx{#1\undefined}
}%
\providecommand \@ifnum [1]{%
 \ifnum #1\expandafter \@firstoftwo
 \else \expandafter \@secondoftwo
 \fi
}%
\providecommand \@ifx [1]{%
 \ifx #1\expandafter \@firstoftwo
 \else \expandafter \@secondoftwo
 \fi
}%
\providecommand \natexlab [1]{#1}%
\providecommand \enquote  [1]{``#1''}%
\providecommand \bibnamefont  [1]{#1}%
\providecommand \bibfnamefont [1]{#1}%
\providecommand \citenamefont [1]{#1}%
\providecommand \href@noop [0]{\@secondoftwo}%
\providecommand \href [0]{\begingroup \@sanitize@url \@href}%
\providecommand \@href[1]{\@@startlink{#1}\@@href}%
\providecommand \@@href[1]{\endgroup#1\@@endlink}%
\providecommand \@sanitize@url [0]{\catcode `\\12\catcode `\$12\catcode `\&12\catcode `\#12\catcode `\^12\catcode `\_12\catcode `\%12\relax}%
\providecommand \@@startlink[1]{}%
\providecommand \@@endlink[0]{}%
\providecommand \url  [0]{\begingroup\@sanitize@url \@url }%
\providecommand \@url [1]{\endgroup\@href {#1}{\urlprefix }}%
\providecommand \urlprefix  [0]{URL }%
\providecommand \Eprint [0]{\href }%
\providecommand \doibase [0]{https://doi.org/}%
\providecommand \selectlanguage [0]{\@gobble}%
\providecommand \bibinfo  [0]{\@secondoftwo}%
\providecommand \bibfield  [0]{\@secondoftwo}%
\providecommand \translation [1]{[#1]}%
\providecommand \BibitemOpen [0]{}%
\providecommand \bibitemStop [0]{}%
\providecommand \bibitemNoStop [0]{.\EOS\space}%
\providecommand \EOS [0]{\spacefactor3000\relax}%
\providecommand \BibitemShut  [1]{\csname bibitem#1\endcsname}%
\let\auto@bib@innerbib\@empty
\bibitem [{\citenamefont {Toda}(2020)}]{toda2020susceptible}%
  \BibitemOpen
  \bibfield  {author} {\bibinfo {author} {\bibfnamefont {A.~A.}\ \bibnamefont {Toda}},\ }\bibfield  {title} {\bibinfo {title} {Susceptible-infected-recovered ({SIR}) dynamics of {C}ovid-19 and economic impact},\ }\href@noop {} {\bibfield  {journal} {\bibinfo  {journal} {Covid Economics}\ ,\ \bibinfo {pages} {43}} (\bibinfo {year} {2020})}\BibitemShut {NoStop}%
\bibitem [{\citenamefont {Osofsky}\ \emph {et~al.}(2020)\citenamefont {Osofsky}, \citenamefont {Osofsky},\ and\ \citenamefont {Mamon}}]{social_impact_covid}%
  \BibitemOpen
  \bibfield  {author} {\bibinfo {author} {\bibfnamefont {J.~D.}\ \bibnamefont {Osofsky}}, \bibinfo {author} {\bibfnamefont {H.~J.}\ \bibnamefont {Osofsky}},\ and\ \bibinfo {author} {\bibfnamefont {L.~Y.}\ \bibnamefont {Mamon}},\ }\bibfield  {title} {\bibinfo {title} {Psychological and social impact of {C}ovid-19.},\ }\href@noop {} {\bibfield  {journal} {\bibinfo  {journal} {Psychological Trauma: Theory, Research, Practice, and Policy}\ }\textbf {\bibinfo {volume} {12}},\ \bibinfo {pages} {468} (\bibinfo {year} {2020})}\BibitemShut {NoStop}%
\bibitem [{\citenamefont {Kermack}\ and\ \citenamefont {McKendrick}(1927)}]{1927_kermack}%
  \BibitemOpen
  \bibfield  {author} {\bibinfo {author} {\bibfnamefont {W.~O.}\ \bibnamefont {Kermack}}\ and\ \bibinfo {author} {\bibfnamefont {A.~G.}\ \bibnamefont {McKendrick}},\ }\bibfield  {title} {\bibinfo {title} {A contribution to the mathematical theory of epidemics},\ }\href@noop {} {\bibfield  {journal} {\bibinfo  {journal} {Proceedings of the royal society of london. Series A, Containing papers of a mathematical and physical character}\ }\textbf {\bibinfo {volume} {115}},\ \bibinfo {pages} {700} (\bibinfo {year} {1927})}\BibitemShut {NoStop}%
\bibitem [{\citenamefont {Hethcote}(2000)}]{mathematics_infectious_diseases}%
  \BibitemOpen
  \bibfield  {author} {\bibinfo {author} {\bibfnamefont {H.~W.}\ \bibnamefont {Hethcote}},\ }\bibfield  {title} {\bibinfo {title} {The mathematics of infectious diseases},\ }\href@noop {} {\bibfield  {journal} {\bibinfo  {journal} {SIAM review}\ }\textbf {\bibinfo {volume} {42}},\ \bibinfo {pages} {599} (\bibinfo {year} {2000})}\BibitemShut {NoStop}%
\bibitem [{\citenamefont {Anderson}\ and\ \citenamefont {May}(1991)}]{anderson1991infectious}%
  \BibitemOpen
  \bibfield  {author} {\bibinfo {author} {\bibfnamefont {R.~M.}\ \bibnamefont {Anderson}}\ and\ \bibinfo {author} {\bibfnamefont {R.~M.}\ \bibnamefont {May}},\ }\href@noop {} {\emph {\bibinfo {title} {Infectious diseases of humans: dynamics and control}}}\ (\bibinfo  {publisher} {Oxford university press},\ \bibinfo {year} {1991})\BibitemShut {NoStop}%
\bibitem [{\citenamefont {Salje}\ \emph {et~al.}(2020)\citenamefont {Salje}, \citenamefont {Tran~Kiem}, \citenamefont {Lefrancq}, \citenamefont {Courtejoie}, \citenamefont {Bosetti}, \citenamefont {Paireau}, \citenamefont {Andronico}, \citenamefont {Hoz{\'e}}, \citenamefont {Richet}, \citenamefont {Dubost} \emph {et~al.}}]{Covid_France_science}%
  \BibitemOpen
  \bibfield  {author} {\bibinfo {author} {\bibfnamefont {H.}~\bibnamefont {Salje}}, \bibinfo {author} {\bibfnamefont {C.}~\bibnamefont {Tran~Kiem}}, \bibinfo {author} {\bibfnamefont {N.}~\bibnamefont {Lefrancq}}, \bibinfo {author} {\bibfnamefont {N.}~\bibnamefont {Courtejoie}}, \bibinfo {author} {\bibfnamefont {P.}~\bibnamefont {Bosetti}}, \bibinfo {author} {\bibfnamefont {J.}~\bibnamefont {Paireau}}, \bibinfo {author} {\bibfnamefont {A.}~\bibnamefont {Andronico}}, \bibinfo {author} {\bibfnamefont {N.}~\bibnamefont {Hoz{\'e}}}, \bibinfo {author} {\bibfnamefont {J.}~\bibnamefont {Richet}}, \bibinfo {author} {\bibfnamefont {C.-L.}\ \bibnamefont {Dubost}}, \emph {et~al.},\ }\bibfield  {title} {\bibinfo {title} {Estimating the burden of {SARS}-{C}o{V}-2 in {F}rance},\ }\href@noop {} {\bibfield  {journal} {\bibinfo  {journal} {Science}\ }\textbf {\bibinfo {volume} {369}},\ \bibinfo {pages} {208} (\bibinfo {year} {2020})}\BibitemShut {NoStop}%
\bibitem [{\citenamefont {Bailey}\ \emph {et~al.}(1975)\citenamefont {Bailey} \emph {et~al.}}]{mathematical_theory_of_infectious_disease}%
  \BibitemOpen
  \bibfield  {author} {\bibinfo {author} {\bibfnamefont {N.~T.}\ \bibnamefont {Bailey}} \emph {et~al.},\ }\href@noop {} {\emph {\bibinfo {title} {The mathematical theory of infectious diseases and its applications}}}\ (\bibinfo  {publisher} {Charles Griffin \& Company Ltd, 5a Crendon Street, High Wycombe, Bucks HP13 6LE.},\ \bibinfo {year} {1975})\BibitemShut {NoStop}%
\bibitem [{\citenamefont {Kendall}(1956)}]{Kendall_1956}%
  \BibitemOpen
  \bibfield  {author} {\bibinfo {author} {\bibfnamefont {D.~G.}\ \bibnamefont {Kendall}},\ }\bibfield  {title} {\bibinfo {title} {Deterministic and stochastic epidemics in closed populations},\ }in\ \href@noop {} {\emph {\bibinfo {booktitle} {Proceedings of the third Berkeley symposium on mathematical statistics and probability}}},\ Vol.~\bibinfo {volume} {4}\ (\bibinfo {organization} {University of California Press Berkeley},\ \bibinfo {year} {1956})\ pp.\ \bibinfo {pages} {149--165}\BibitemShut {NoStop}%
\bibitem [{\citenamefont {Harko}\ \emph {et~al.}(2014)\citenamefont {Harko}, \citenamefont {Lobo},\ and\ \citenamefont {Mak}}]{Exact_analytical_SIR_2014}%
  \BibitemOpen
  \bibfield  {author} {\bibinfo {author} {\bibfnamefont {T.}~\bibnamefont {Harko}}, \bibinfo {author} {\bibfnamefont {F.~S.}\ \bibnamefont {Lobo}},\ and\ \bibinfo {author} {\bibfnamefont {M.}~\bibnamefont {Mak}},\ }\bibfield  {title} {\bibinfo {title} {Exact analytical solutions of the susceptible-infected-recovered ({SIR}) epidemic model and of the {SIR} model with equal death and birth rates},\ }\href@noop {} {\bibfield  {journal} {\bibinfo  {journal} {Applied Mathematics and Computation}\ }\textbf {\bibinfo {volume} {236}},\ \bibinfo {pages} {184} (\bibinfo {year} {2014})}\BibitemShut {NoStop}%
\bibitem [{\citenamefont {Prodanov}(2022)}]{Prodanov2022_analytical_SIR}%
  \BibitemOpen
  \bibfield  {author} {\bibinfo {author} {\bibfnamefont {D.}~\bibnamefont {Prodanov}},\ }\bibfield  {title} {\bibinfo {title} {Analytical solutions and parameter estimation of the {SIR} epidemic model},\ }\href@noop {} {\bibfield  {journal} {\bibinfo  {journal} {Mathematical Analysis of Infectious Diseases}\ ,\ \bibinfo {pages} {163}} (\bibinfo {year} {2022})}\BibitemShut {NoStop}%
\bibitem [{\citenamefont {Carvalho}\ and\ \citenamefont {Gon{\c{c}}alves}(2021)}]{carvalho2021analytical}%
  \BibitemOpen
  \bibfield  {author} {\bibinfo {author} {\bibfnamefont {A.~M.}\ \bibnamefont {Carvalho}}\ and\ \bibinfo {author} {\bibfnamefont {S.}~\bibnamefont {Gon{\c{c}}alves}},\ }\bibfield  {title} {\bibinfo {title} {An analytical solution for the {K}ermack--{McK}endrick model},\ }\href@noop {} {\bibfield  {journal} {\bibinfo  {journal} {Phys. A: Stat. Mech.}\ }\textbf {\bibinfo {volume} {566}},\ \bibinfo {pages} {125659} (\bibinfo {year} {2021})}\BibitemShut {NoStop}%
\bibitem [{\citenamefont {Li}\ and\ \citenamefont {Guo}(2017)}]{SIRC}%
  \BibitemOpen
  \bibfield  {author} {\bibinfo {author} {\bibfnamefont {H.}~\bibnamefont {Li}}\ and\ \bibinfo {author} {\bibfnamefont {S.}~\bibnamefont {Guo}},\ }\bibfield  {title} {\bibinfo {title} {Dynamics of a {SIRC} epidemiological model},\ }\href@noop {} {\bibfield  {journal} {\bibinfo  {journal} {Electron. J. Differ. Equ.}\ } (\bibinfo {year} {2017})}\BibitemShut {NoStop}%
\bibitem [{\citenamefont {Li}\ \emph {et~al.}(2001)\citenamefont {Li}, \citenamefont {Smith},\ and\ \citenamefont {Wang}}]{SEIR}%
  \BibitemOpen
  \bibfield  {author} {\bibinfo {author} {\bibfnamefont {M.~Y.}\ \bibnamefont {Li}}, \bibinfo {author} {\bibfnamefont {H.~L.}\ \bibnamefont {Smith}},\ and\ \bibinfo {author} {\bibfnamefont {L.}~\bibnamefont {Wang}},\ }\bibfield  {title} {\bibinfo {title} {Global dynamics of an {SEIR} epidemic model with vertical transmission},\ }\href@noop {} {\bibfield  {journal} {\bibinfo  {journal} {SIAM J. Appl. Math.}\ }\textbf {\bibinfo {volume} {62}},\ \bibinfo {pages} {58} (\bibinfo {year} {2001})}\BibitemShut {NoStop}%
\bibitem [{\citenamefont {Sen}\ and\ \citenamefont {Sen}(2021)}]{SIRD}%
  \BibitemOpen
  \bibfield  {author} {\bibinfo {author} {\bibfnamefont {D.}~\bibnamefont {Sen}}\ and\ \bibinfo {author} {\bibfnamefont {D.}~\bibnamefont {Sen}},\ }\bibfield  {title} {\bibinfo {title} {Use of a modified {SIRD} model to analyze {C}ovid-19 data},\ }\href@noop {} {\bibfield  {journal} {\bibinfo  {journal} {Ind. Eng. Chem. Res.}\ }\textbf {\bibinfo {volume} {60}},\ \bibinfo {pages} {4251} (\bibinfo {year} {2021})}\BibitemShut {NoStop}%
\bibitem [{\citenamefont {Gao}\ \emph {et~al.}(2007)\citenamefont {Gao}, \citenamefont {Teng}, \citenamefont {Nieto}, \citenamefont {Torres} \emph {et~al.}}]{SIRV}%
  \BibitemOpen
  \bibfield  {author} {\bibinfo {author} {\bibfnamefont {S.}~\bibnamefont {Gao}}, \bibinfo {author} {\bibfnamefont {Z.}~\bibnamefont {Teng}}, \bibinfo {author} {\bibfnamefont {J.~J.}\ \bibnamefont {Nieto}}, \bibinfo {author} {\bibfnamefont {A.}~\bibnamefont {Torres}}, \emph {et~al.},\ }\bibfield  {title} {\bibinfo {title} {Analysis of an {SIR} epidemic model with pulse vaccination and distributed time delay},\ }\href@noop {} {\bibfield  {journal} {\bibinfo  {journal} {BioMed Research International}\ }\textbf {\bibinfo {volume} {2007}} (\bibinfo {year} {2007})}\BibitemShut {NoStop}%
\bibitem [{\citenamefont {Kozyreff}(2022)}]{Asymptotic_SIR_SEIR}%
  \BibitemOpen
  \bibfield  {author} {\bibinfo {author} {\bibfnamefont {G.}~\bibnamefont {Kozyreff}},\ }\bibfield  {title} {\bibinfo {title} {Asymptotic solutions of the {SIR} and {SEIR} models well above the epidemic threshold},\ }\href@noop {} {\bibfield  {journal} {\bibinfo  {journal} {IMA Journal of Applied Mathematics}\ }\textbf {\bibinfo {volume} {87}},\ \bibinfo {pages} {521} (\bibinfo {year} {2022})}\BibitemShut {NoStop}%
\bibitem [{\citenamefont {Fumanelli}\ \emph {et~al.}(2012)\citenamefont {Fumanelli}, \citenamefont {Ajelli}, \citenamefont {Manfredi}, \citenamefont {Vespignani},\ and\ \citenamefont {Merler}}]{Inferring_social_structure}%
  \BibitemOpen
  \bibfield  {author} {\bibinfo {author} {\bibfnamefont {L.}~\bibnamefont {Fumanelli}}, \bibinfo {author} {\bibfnamefont {M.}~\bibnamefont {Ajelli}}, \bibinfo {author} {\bibfnamefont {P.}~\bibnamefont {Manfredi}}, \bibinfo {author} {\bibfnamefont {A.}~\bibnamefont {Vespignani}},\ and\ \bibinfo {author} {\bibfnamefont {S.}~\bibnamefont {Merler}},\ }\bibfield  {title} {\bibinfo {title} {Inferring the structure of social contacts from demographic data in the analysis of infectious diseases spread},\ }\href@noop {} {\bibfield  {journal} {\bibinfo  {journal} {PLoS Comput Biol}\ }\textbf {\bibinfo {volume} {8}} (\bibinfo {year} {2012})}\BibitemShut {NoStop}%
\bibitem [{\citenamefont {Mistry}\ \emph {et~al.}(2021)\citenamefont {Mistry}, \citenamefont {Litvinova}, \citenamefont {Pastore~y Piontti}, \citenamefont {Chinazzi}, \citenamefont {Fumanelli}, \citenamefont {Gomes}, \citenamefont {Haque}, \citenamefont {Liu}, \citenamefont {Mu}, \citenamefont {Xiong} \emph {et~al.}}]{mistry2020inferring}%
  \BibitemOpen
  \bibfield  {author} {\bibinfo {author} {\bibfnamefont {D.}~\bibnamefont {Mistry}}, \bibinfo {author} {\bibfnamefont {M.}~\bibnamefont {Litvinova}}, \bibinfo {author} {\bibfnamefont {A.}~\bibnamefont {Pastore~y Piontti}}, \bibinfo {author} {\bibfnamefont {M.}~\bibnamefont {Chinazzi}}, \bibinfo {author} {\bibfnamefont {L.}~\bibnamefont {Fumanelli}}, \bibinfo {author} {\bibfnamefont {M.~F.}\ \bibnamefont {Gomes}}, \bibinfo {author} {\bibfnamefont {S.~A.}\ \bibnamefont {Haque}}, \bibinfo {author} {\bibfnamefont {Q.-H.}\ \bibnamefont {Liu}}, \bibinfo {author} {\bibfnamefont {K.}~\bibnamefont {Mu}}, \bibinfo {author} {\bibfnamefont {X.}~\bibnamefont {Xiong}}, \emph {et~al.},\ }\bibfield  {title} {\bibinfo {title} {Inferring high-resolution human mixing patterns for disease modeling},\ }\href@noop {} {\bibfield  {journal} {\bibinfo  {journal} {Nature comm.}\ }\textbf {\bibinfo {volume} {12}},\ \bibinfo {pages} {323} (\bibinfo {year} {2021})}\BibitemShut {NoStop}%
\bibitem [{\citenamefont {Ferguson}\ \emph {et~al.}(2020)\citenamefont {Ferguson}, \citenamefont {Laydon}, \citenamefont {Nedjati~Gilani}, \citenamefont {Imai}, \citenamefont {Ainslie}, \citenamefont {Baguelin}, \citenamefont {Bhatia}, \citenamefont {Boonyasiri}, \citenamefont {Cucunuba~Perez}, \citenamefont {Cuomo-Dannenburg} \emph {et~al.}}]{Impact_NPI_Imperial}%
  \BibitemOpen
  \bibfield  {author} {\bibinfo {author} {\bibfnamefont {N.}~\bibnamefont {Ferguson}}, \bibinfo {author} {\bibfnamefont {D.}~\bibnamefont {Laydon}}, \bibinfo {author} {\bibfnamefont {G.}~\bibnamefont {Nedjati~Gilani}}, \bibinfo {author} {\bibfnamefont {N.}~\bibnamefont {Imai}}, \bibinfo {author} {\bibfnamefont {K.}~\bibnamefont {Ainslie}}, \bibinfo {author} {\bibfnamefont {M.}~\bibnamefont {Baguelin}}, \bibinfo {author} {\bibfnamefont {S.}~\bibnamefont {Bhatia}}, \bibinfo {author} {\bibfnamefont {A.}~\bibnamefont {Boonyasiri}}, \bibinfo {author} {\bibfnamefont {Z.}~\bibnamefont {Cucunuba~Perez}}, \bibinfo {author} {\bibfnamefont {G.}~\bibnamefont {Cuomo-Dannenburg}}, \emph {et~al.},\ }\bibfield  {title} {\bibinfo {title} {Report 9: Impact of non-pharmaceutical interventions ({NPI}s) to reduce {C}ovid19 mortality and healthcare demand},\ }\href@noop {} {\bibfield  {journal} {\bibinfo  {journal} {Imperial College London COVID-19}\ } (\bibinfo {year} {2020})}\BibitemShut {NoStop}%
\bibitem [{\citenamefont {Bremaud}\ and\ \citenamefont {Ullmo}(2022)}]{Bremaud_Ullmo_PhysRevE}%
  \BibitemOpen
  \bibfield  {author} {\bibinfo {author} {\bibfnamefont {L.}~\bibnamefont {Bremaud}}\ and\ \bibinfo {author} {\bibfnamefont {D.}~\bibnamefont {Ullmo}},\ }\bibfield  {title} {\bibinfo {title} {Social structure description of epidemic propagation with a mean-field game paradigm},\ }\href@noop {} {\bibfield  {journal} {\bibinfo  {journal} {Phys. Rev. E}\ }\textbf {\bibinfo {volume} {106}},\ \bibinfo {pages} {L062301} (\bibinfo {year} {2022})}\BibitemShut {NoStop}%
\bibitem [{\citenamefont {Watts}\ and\ \citenamefont {Strogatz}(1998)}]{Watts1998-small_world}%
  \BibitemOpen
  \bibfield  {author} {\bibinfo {author} {\bibfnamefont {D.~J.}\ \bibnamefont {Watts}}\ and\ \bibinfo {author} {\bibfnamefont {S.~H.}\ \bibnamefont {Strogatz}},\ }\bibfield  {title} {\bibinfo {title} {Collective dynamics of ‘small-world’networks},\ }\href@noop {} {\bibfield  {journal} {\bibinfo  {journal} {Nature}\ }\textbf {\bibinfo {volume} {393}},\ \bibinfo {pages} {440} (\bibinfo {year} {1998})}\BibitemShut {NoStop}%
\bibitem [{\citenamefont {Santos}\ \emph {et~al.}(2005)\citenamefont {Santos}, \citenamefont {Rodrigues},\ and\ \citenamefont {Pacheco}}]{PhysRevE_cooperation_small_world_networks}%
  \BibitemOpen
  \bibfield  {author} {\bibinfo {author} {\bibfnamefont {F.~C.}\ \bibnamefont {Santos}}, \bibinfo {author} {\bibfnamefont {J.~F.}\ \bibnamefont {Rodrigues}},\ and\ \bibinfo {author} {\bibfnamefont {J.~M.}\ \bibnamefont {Pacheco}},\ }\bibfield  {title} {\bibinfo {title} {Epidemic spreading and cooperation dynamics on homogeneous small-world networks},\ }\href@noop {} {\bibfield  {journal} {\bibinfo  {journal} {Phys. Rev. E}\ }\textbf {\bibinfo {volume} {72}},\ \bibinfo {pages} {056128} (\bibinfo {year} {2005})}\BibitemShut {NoStop}%
\bibitem [{\citenamefont {Silva}\ \emph {et~al.}(2020)\citenamefont {Silva}, \citenamefont {Rodrigues},\ and\ \citenamefont {Ferreira}}]{PhysRevE_pair_quenched_MFT_threshold_SIS}%
  \BibitemOpen
  \bibfield  {author} {\bibinfo {author} {\bibfnamefont {D.~H.}\ \bibnamefont {Silva}}, \bibinfo {author} {\bibfnamefont {F.~A.}\ \bibnamefont {Rodrigues}},\ and\ \bibinfo {author} {\bibfnamefont {S.~C.}\ \bibnamefont {Ferreira}},\ }\bibfield  {title} {\bibinfo {title} {High prevalence regimes in the pair-quenched mean-field theory for the susceptible-infected-susceptible model on networks},\ }\href@noop {} {\bibfield  {journal} {\bibinfo  {journal} {Phys. Rev. E}\ }\textbf {\bibinfo {volume} {102}},\ \bibinfo {pages} {012313} (\bibinfo {year} {2020})}\BibitemShut {NoStop}%
\bibitem [{\citenamefont {Mizutaka}\ \emph {et~al.}(2022)\citenamefont {Mizutaka}, \citenamefont {Mori},\ and\ \citenamefont {Hasegawa}}]{PhysRevE_synergistic_epidemic}%
  \BibitemOpen
  \bibfield  {author} {\bibinfo {author} {\bibfnamefont {S.}~\bibnamefont {Mizutaka}}, \bibinfo {author} {\bibfnamefont {K.}~\bibnamefont {Mori}},\ and\ \bibinfo {author} {\bibfnamefont {T.}~\bibnamefont {Hasegawa}},\ }\bibfield  {title} {\bibinfo {title} {Synergistic epidemic spreading in correlated networks},\ }\href@noop {} {\bibfield  {journal} {\bibinfo  {journal} {Phys. Rev. E}\ }\textbf {\bibinfo {volume} {106}},\ \bibinfo {pages} {034305} (\bibinfo {year} {2022})}\BibitemShut {NoStop}%
\bibitem [{\citenamefont {Barth{\'e}lemy}\ \emph {et~al.}(2005)\citenamefont {Barth{\'e}lemy}, \citenamefont {Barrat}, \citenamefont {Pastor-Satorras},\ and\ \citenamefont {Vespignani}}]{Barthelemy2005_networks}%
  \BibitemOpen
  \bibfield  {author} {\bibinfo {author} {\bibfnamefont {M.}~\bibnamefont {Barth{\'e}lemy}}, \bibinfo {author} {\bibfnamefont {A.}~\bibnamefont {Barrat}}, \bibinfo {author} {\bibfnamefont {R.}~\bibnamefont {Pastor-Satorras}},\ and\ \bibinfo {author} {\bibfnamefont {A.}~\bibnamefont {Vespignani}},\ }\bibfield  {title} {\bibinfo {title} {Dynamical patterns of epidemic outbreaks in complex heterogeneous networks},\ }\href@noop {} {\bibfield  {journal} {\bibinfo  {journal} {Journal of theoretical biology}\ }\textbf {\bibinfo {volume} {235}},\ \bibinfo {pages} {275} (\bibinfo {year} {2005})}\BibitemShut {NoStop}%
\bibitem [{\citenamefont {Hu}\ \emph {et~al.}(2015)\citenamefont {Hu}, \citenamefont {Min},\ and\ \citenamefont {Kuang}}]{dynamics_spreading_homogeneous_heterogeneous}%
  \BibitemOpen
  \bibfield  {author} {\bibinfo {author} {\bibfnamefont {Y.}~\bibnamefont {Hu}}, \bibinfo {author} {\bibfnamefont {L.}~\bibnamefont {Min}},\ and\ \bibinfo {author} {\bibfnamefont {Y.}~\bibnamefont {Kuang}},\ }\bibfield  {title} {\bibinfo {title} {Modeling the dynamics of epidemic spreading on homogenous and heterogeneous networks},\ }\href@noop {} {\bibfield  {journal} {\bibinfo  {journal} {Applicable Analysis}\ }\textbf {\bibinfo {volume} {94}},\ \bibinfo {pages} {2308} (\bibinfo {year} {2015})}\BibitemShut {NoStop}%
\bibitem [{\citenamefont {Sahneh}\ and\ \citenamefont {Scoglio}(2011)}]{Epidemc_spread_networks}%
  \BibitemOpen
  \bibfield  {author} {\bibinfo {author} {\bibfnamefont {F.~D.}\ \bibnamefont {Sahneh}}\ and\ \bibinfo {author} {\bibfnamefont {C.}~\bibnamefont {Scoglio}},\ }\bibfield  {title} {\bibinfo {title} {Epidemic spread in human networks},\ }in\ \href@noop {} {\emph {\bibinfo {booktitle} {2011 50th IEEE Conference on Decision and Control and European Control Conference}}}\ (\bibinfo {organization} {IEEE},\ \bibinfo {year} {2011})\ pp.\ \bibinfo {pages} {3008--3013}\BibitemShut {NoStop}%
\bibitem [{\citenamefont {Newman}(2002)}]{newman2002spread}%
  \BibitemOpen
  \bibfield  {author} {\bibinfo {author} {\bibfnamefont {M.~E.}\ \bibnamefont {Newman}},\ }\bibfield  {title} {\bibinfo {title} {Spread of epidemic disease on networks},\ }\href@noop {} {\bibfield  {journal} {\bibinfo  {journal} {Phys. Rev. E}\ }\textbf {\bibinfo {volume} {66}},\ \bibinfo {pages} {016128} (\bibinfo {year} {2002})}\BibitemShut {NoStop}%
\bibitem [{\citenamefont {Di~Domenico}(2022)}]{didomenico_thesis}%
  \BibitemOpen
  \bibfield  {author} {\bibinfo {author} {\bibfnamefont {L.}~\bibnamefont {Di~Domenico}},\ }\emph {\bibinfo {title} {Data-driven modeling of COVID-19 spread in France to inform pandemic response}},\ \href@noop {} {Ph.D. thesis},\ \bibinfo  {school} {Sorbonne Universit{\'e}} (\bibinfo {year} {2022})\BibitemShut {NoStop}%
\bibitem [{\citenamefont {Pastor-Satorras}\ \emph {et~al.}(2015)\citenamefont {Pastor-Satorras}, \citenamefont {Castellano}, \citenamefont {Van~Mieghem},\ and\ \citenamefont {Vespignani}}]{Epidemic_processes_complex_networks}%
  \BibitemOpen
  \bibfield  {author} {\bibinfo {author} {\bibfnamefont {R.}~\bibnamefont {Pastor-Satorras}}, \bibinfo {author} {\bibfnamefont {C.}~\bibnamefont {Castellano}}, \bibinfo {author} {\bibfnamefont {P.}~\bibnamefont {Van~Mieghem}},\ and\ \bibinfo {author} {\bibfnamefont {A.}~\bibnamefont {Vespignani}},\ }\bibfield  {title} {\bibinfo {title} {Epidemic processes in complex networks},\ }\href@noop {} {\bibfield  {journal} {\bibinfo  {journal} {Rev. Mod. Phys.}\ }\textbf {\bibinfo {volume} {87}},\ \bibinfo {pages} {925} (\bibinfo {year} {2015})}\BibitemShut {NoStop}%
\bibitem [{\citenamefont {Wang}\ \emph {et~al.}(2017)\citenamefont {Wang}, \citenamefont {Tang}, \citenamefont {Stanley},\ and\ \citenamefont {Braunstein}}]{Wang_2017_unification}%
  \BibitemOpen
  \bibfield  {author} {\bibinfo {author} {\bibfnamefont {W.}~\bibnamefont {Wang}}, \bibinfo {author} {\bibfnamefont {M.}~\bibnamefont {Tang}}, \bibinfo {author} {\bibfnamefont {H.~E.}\ \bibnamefont {Stanley}},\ and\ \bibinfo {author} {\bibfnamefont {L.~A.}\ \bibnamefont {Braunstein}},\ }\bibfield  {title} {\bibinfo {title} {Unification of theoretical approaches for epidemic spreading on complex networks},\ }\href@noop {} {\bibfield  {journal} {\bibinfo  {journal} {Rep. Prog. Phys.}\ }\textbf {\bibinfo {volume} {80}},\ \bibinfo {pages} {036603} (\bibinfo {year} {2017})}\BibitemShut {NoStop}%
\bibitem [{\citenamefont {Allen}(2008)}]{stochastic_models}%
  \BibitemOpen
  \bibfield  {author} {\bibinfo {author} {\bibfnamefont {L.~J.}\ \bibnamefont {Allen}},\ }\bibfield  {title} {\bibinfo {title} {An introduction to stochastic epidemic models},\ }in\ \href@noop {} {\emph {\bibinfo {booktitle} {Mathematical epidemiology}}}\ (\bibinfo  {publisher} {Springer},\ \bibinfo {year} {2008})\ pp.\ \bibinfo {pages} {81--130}\BibitemShut {NoStop}%
\bibitem [{\citenamefont {Simon}\ and\ \citenamefont {Kiss}(2016)}]{compact_pairwise_heteregogeneous}%
  \BibitemOpen
  \bibfield  {author} {\bibinfo {author} {\bibfnamefont {P.~L.}\ \bibnamefont {Simon}}\ and\ \bibinfo {author} {\bibfnamefont {I.~Z.}\ \bibnamefont {Kiss}},\ }\bibfield  {title} {\bibinfo {title} {Super compact pairwise model for {SIS} epidemic on heterogeneous networks},\ }\href@noop {} {\bibfield  {journal} {\bibinfo  {journal} {J. Complex Netw.}\ }\textbf {\bibinfo {volume} {4}},\ \bibinfo {pages} {187} (\bibinfo {year} {2016})}\BibitemShut {NoStop}%
\bibitem [{\citenamefont {Wormald}(1981)}]{wormald1981asymptotic}%
  \BibitemOpen
  \bibfield  {author} {\bibinfo {author} {\bibfnamefont {N.~C.}\ \bibnamefont {Wormald}},\ }\bibfield  {title} {\bibinfo {title} {The asymptotic connectivity of labelled regular graphs},\ }\href@noop {} {\bibfield  {journal} {\bibinfo  {journal} {J. Comb. Theory. Ser. B}\ }\textbf {\bibinfo {volume} {31}},\ \bibinfo {pages} {156} (\bibinfo {year} {1981})}\BibitemShut {NoStop}%
\bibitem [{\citenamefont {Dorogovtsev}\ \emph {et~al.}(2003)\citenamefont {Dorogovtsev}, \citenamefont {Goltsev}, \citenamefont {Mendes},\ and\ \citenamefont {Samukhin}}]{dorogovtsev2003spectra}%
  \BibitemOpen
  \bibfield  {author} {\bibinfo {author} {\bibfnamefont {S.~N.}\ \bibnamefont {Dorogovtsev}}, \bibinfo {author} {\bibfnamefont {A.~V.}\ \bibnamefont {Goltsev}}, \bibinfo {author} {\bibfnamefont {J.~F.}\ \bibnamefont {Mendes}},\ and\ \bibinfo {author} {\bibfnamefont {A.~N.}\ \bibnamefont {Samukhin}},\ }\bibfield  {title} {\bibinfo {title} {Spectra of complex networks},\ }\href@noop {} {\bibfield  {journal} {\bibinfo  {journal} {Phys. Rev. E}\ }\textbf {\bibinfo {volume} {68}},\ \bibinfo {pages} {046109} (\bibinfo {year} {2003})}\BibitemShut {NoStop}%
\bibitem [{\citenamefont {Goirand}\ \emph {et~al.}(2021)\citenamefont {Goirand}, \citenamefont {Georgeot}, \citenamefont {Giraud},\ and\ \citenamefont {Lorthois}}]{goirand2021network}%
  \BibitemOpen
  \bibfield  {author} {\bibinfo {author} {\bibfnamefont {F.}~\bibnamefont {Goirand}}, \bibinfo {author} {\bibfnamefont {B.}~\bibnamefont {Georgeot}}, \bibinfo {author} {\bibfnamefont {O.}~\bibnamefont {Giraud}},\ and\ \bibinfo {author} {\bibfnamefont {S.}~\bibnamefont {Lorthois}},\ }\bibfield  {title} {\bibinfo {title} {Network community structure and resilience to localized damage: application to brain microcirculation},\ }\href@noop {} {\bibfield  {journal} {\bibinfo  {journal} {Brain Multiphysics}\ }\textbf {\bibinfo {volume} {2}},\ \bibinfo {pages} {100028} (\bibinfo {year} {2021})}\BibitemShut {NoStop}%
\bibitem [{\citenamefont {Lambert}(1758)}]{lambert1758observationes}%
  \BibitemOpen
  \bibfield  {author} {\bibinfo {author} {\bibfnamefont {J.~H.}\ \bibnamefont {Lambert}},\ }\bibfield  {title} {\bibinfo {title} {Observationes variae in mathesin puram},\ }\href@noop {} {\bibfield  {journal} {\bibinfo  {journal} {Acta Helvetica}\ }\textbf {\bibinfo {volume} {3}},\ \bibinfo {pages} {128} (\bibinfo {year} {1758})}\BibitemShut {NoStop}%
\bibitem [{\citenamefont {Corless}\ \emph {et~al.}(1996)\citenamefont {Corless}, \citenamefont {Gonnet}, \citenamefont {Hare}, \citenamefont {Jeffrey},\ and\ \citenamefont {Knuth}}]{On_LambertW_function}%
  \BibitemOpen
  \bibfield  {author} {\bibinfo {author} {\bibfnamefont {R.~M.}\ \bibnamefont {Corless}}, \bibinfo {author} {\bibfnamefont {G.~H.}\ \bibnamefont {Gonnet}}, \bibinfo {author} {\bibfnamefont {D.~E.}\ \bibnamefont {Hare}}, \bibinfo {author} {\bibfnamefont {D.~J.}\ \bibnamefont {Jeffrey}},\ and\ \bibinfo {author} {\bibfnamefont {D.~E.}\ \bibnamefont {Knuth}},\ }\bibfield  {title} {\bibinfo {title} {On the {L}ambert {W} function},\ }\href@noop {} {\bibfield  {journal} {\bibinfo  {journal} {Adv. Comput. Math.}\ }\textbf {\bibinfo {volume} {5}},\ \bibinfo {pages} {329} (\bibinfo {year} {1996})}\BibitemShut {NoStop}%
\bibitem [{\citenamefont {Euler}(1783)}]{euler1783serie}%
  \BibitemOpen
  \bibfield  {author} {\bibinfo {author} {\bibfnamefont {L.}~\bibnamefont {Euler}},\ }\bibfield  {title} {\bibinfo {title} {De serie lambertina plurimisque eius insignibus proprietatibus},\ }\href@noop {} {\bibfield  {journal} {\bibinfo  {journal} {Acta Academiae scientiarum imperialis petropolitanae}\ ,\ \bibinfo {pages} {29}} (\bibinfo {year} {1783})}\BibitemShut {NoStop}%
\bibitem [{\citenamefont {Belki{\'c}}(2019)}]{belkic2019all}%
  \BibitemOpen
  \bibfield  {author} {\bibinfo {author} {\bibfnamefont {D.}~\bibnamefont {Belki{\'c}}},\ }\bibfield  {title} {\bibinfo {title} {All the trinomial roots, their powers and logarithms from the {L}ambert series, bell polynomials and {F}ox-{W}right function: Illustration for genome multiplicity in survival of irradiated cells},\ }\href@noop {} {\bibfield  {journal} {\bibinfo  {journal} {Journal of Mathematical Chemistry}\ }\textbf {\bibinfo {volume} {57}},\ \bibinfo {pages} {59} (\bibinfo {year} {2019})}\BibitemShut {NoStop}%
\bibitem [{\citenamefont {Glasser}(2000)}]{glasser2000hypergeometric}%
  \BibitemOpen
  \bibfield  {author} {\bibinfo {author} {\bibfnamefont {M.}~\bibnamefont {Glasser}},\ }\bibfield  {title} {\bibinfo {title} {Hypergeometric functions and the trinomial equation},\ }\href@noop {} {\bibfield  {journal} {\bibinfo  {journal} {J. Comput. Appl. Math.}\ }\textbf {\bibinfo {volume} {118}},\ \bibinfo {pages} {169} (\bibinfo {year} {2000})}\BibitemShut {NoStop}%
\end{thebibliography}%

\end{document}